\documentclass[iop]{emulateapj}
\usepackage{epstopdf}

\newcommand{\be}{\begin{eqnarray}}
\newcommand{\ee}{\end{eqnarray}}
\newcommand{\lp}{\left(}
\newcommand{\rp}{\right)}

\newcommand{\E}[1]{\times10^{#1}}
\newcommand{\smpy}{ \ M_\odot \ {\rm yr}^{-1}}
\newcommand{\msol}{ \ M_\odot }
\newcommand{\commentOut}[1]{}
\newcommand{\bi}{\begin{itemize}}
\newcommand{\ei}{\end{itemize}}
\newcommand{\cgsd}{ {\rm \ g \ cm^{-3}}}
\newcommand{\cgsv}{ {\rm \ cm \ s^{-1}}}

\shorttitle{CONVERGING SHOCK IGNITION OF DETONATIONS IN WHITE DWARFS}
\shortauthors{SHEN \& BILDSTEN}
\slugcomment{Accepted for publication in \apj}

\begin{document}


\title{The Ignition of Carbon Detonations via Converging Shock Waves in White Dwarfs}

\author{Ken J. Shen\altaffilmark{1,2}}
\altaffiltext{1}{Department of Astronomy and Theoretical Astrophysics Center, University of California, Berkeley, CA 94720, USA; kenshen@astro.berkeley.edu.}
\altaffiltext{2}{Einstein Fellow.}

\author{Lars Bildsten\altaffilmark{3}}
\altaffiltext{3}{Kavli Institute for Theoretical Physics and Department of Physics, Kohn Hall, University of California, Santa Barbara, CA 93106; bildsten@kitp.ucsb.edu.}


\begin{abstract}

The progenitor channel responsible for the majority of Type Ia supernovae is still uncertain.  One emergent scenario involves the detonation of a He-rich layer surrounding a C/O white dwarf, which sends a shock wave into the core.  The quasi-spherical shock wave converges and strengthens at an off-center location, forming a second, C-burning, detonation that disrupts the whole star.  In this paper, we examine this second detonation of the double detonation scenario using a combination of analytic and numeric techniques.  We perform a spatially resolved study of the imploding shock wave and outgoing detonation and calculate the critical imploding shock strengths needed to achieve a core C detonation.  We find that He detonations in recent two-dimensional simulations yield converging shock waves that are strong enough to ignite C detonations in high-mass C/O cores, with the caveat that a truly robust answer requires multi-dimensional detonation initiation calculations.  We also find that convergence-driven detonations in low-mass C/O cores and in O/Ne cores are harder to achieve and are perhaps unrealized in standard binary evolution.

\end{abstract}



\section{Introduction}
\label{sec:intro}

Despite decades of theory and observations, the nature of Type Ia supernova (SN Ia) progenitors remains a mystery.  Until recently, the evolutionary scenario thought to be responsible for the bulk of SNe Ia was the ``single degenerate scenario,'' in which a C/O white dwarf (WD) accretes H-rich matter from a donor and ignites C in its core as it approaches the Chandrasekhar mass \citep{wi73,nomo82a}.  However, recent work has revealed cracks in this scenario from a variety of angles (e.g., \citealt{leon07,nomo07,sb07,rbf09,kase10,li11,sp12}).  Studies of the ``double degenerate scenario'' \citep{it84,webb84}, which involves the growth of a C/O WD to the Chandrasekhar mass via the merger of two C/O WDs, have shown that it instead results in relatively quiescent C-burning rather than the violent deflagration or detonation necessary for a SN Ia \citep{ni85,sn98,ypr07,shen12,schw12}.

Our focus in this work is on the ``double detonation scenario,'' in which a detonation in a He shell surrounding a C/O WD sends a shock wave into the C/O core.  As this quasi-spherical shock wave converges towards a focal point, it strengthens and subsequently ignites a C-burning detonation (e.g., \citealt{livn90}).  The double detonation scenario neatly explains why shock interaction of the ejecta with large companions, significant H, and pre- and post-explosion companions are not detected in typical SNe Ia, and the scenario can provide a very good match to the observed SN Ia delay time distribution \citep{maoz11,ruit11}.  However, while the double detonation scenario was first invoked several decades ago, no study has adequately resolved the ignition of the second detonation.  This is due to the large disparity between the scale of the WD ($10^8-10^9$ cm) and the lengthscales of C detonations ($0.01-1$ cm) at the relevant densities of $\rho = 10^7-10^8 \cgsd$.  This $\sim 10$ order of magnitude difference in length highlights the computational challenge of resolving the core detonation ignition in a full star simulation.  As a result, studies that track the progress of the He shell detonation and the shock wave through the entire C/O core, which have a minimum resolution of $\sim 10^6$ cm \citep{fhr07,fink10,sim12,moll13a}, typically assume that if the minimum resolution element is compressed to high densities and temperatures, a C detonation is inevitable.  However, this assumption has not yet been properly tested.

In this paper, we narrow our attention to an initially constant density volume surrounding the focal point of the converging shock wave in order to resolve the formation of the core detonation.  We begin in Section \ref{sec:evolscen} by outlining the evolutionary pathways that can lead to double detonations.  In Section \ref{sec:analytics}, we describe analytic results for planar detonations and spherically imploding shock waves.  We use the numerical reactive hydrodynamics code FLASH \citep{fryx00} to follow the imploding shock wave and the ignition of the detonation in Section \ref{sec:numerics}, and we find the critical spherical imploding shock strengths needed to achieve propagating spherical detonations.  Our results support claims (e.g., \citealt{fhr07,fink10,sim10,ruit11}) that He shell detonations yield converging shocks that are strong enough to ignite detonations in high-mass C/O cores.  However, detonations in smaller C/O cores are harder to ignite, and O/Ne cores appear to be prohibitively difficult to detonate.  We summarize our work and conclude in Section \ref{sec:conc}.


\section{Evolution of double detonation progenitors}
\label{sec:evolscen}

The double detonation scenario was first considered in the context of prolonged mass transfer of He onto C/O WDs at accretion rates $ \sim 10^{-8} \smpy$ \citep{nomo82b,wtw86,livn90,lg90,lg91,ww94,la95}.  The mass donor in these early works was a $\simeq 0.5 \msol$ He-burning sdB/sdO star, which yields relatively large ($ \gtrsim0.1 \msol)$ He envelopes on the C/O accretor prior to He ignition.  When convective shell burning progresses in these large envelopes, convective eddies become inefficient at carrying away the energy released in the thin burning layer, and a He detonation may develop \citep{taam80a,taam80b}.  The shock wave sent into the C/O core may be large enough to immediately ignite a C-detonation upon encountering the C-rich material, sometimes referred to as an ``edge-lit'' detonation, or may converge near the center of the core and form a detonation there \citep{nomo82b,livn90}.  In this paper, we will restrict our analysis to the latter, convergence-driven, channel.

While the energetics and nucleosynthesis from the C/O core detonation roughly matched SN Ia light curves, more detailed spectral comparisons failed because of the large amount of iron-group elements (IGEs) produced in the thick He shell detonation \citep{hk96,nuge97}.  Furthermore, the predicted binary population synthesis rate of explosions from such an evolutionary channel is too low to account for the bulk of SNe Ia, particularly in old stellar populations \citep{ruit11}.  It is also possible that at these relatively high densities, the dynamical He-burning progresses as a He deflagration instead of a detonation \citep{wk11}, potentially yielding the newly discovered classes of SN 2002cx-like / SNe Iax objects \citep{li03,phil07,fole09,fole13a} or Ca-rich / O-poor transients \citep{pere10,kasl12}.

In more recent years, the possibility of He shell detonations in systems with dynamically stable mass transfer from a He WD donor was considered \citep{bild07,sb09b,shen10,kbs12}.  Because the resulting accretion rates are higher, the accumulated He shells at the onset of He-burning are $10-100$ times less massive than in the non-degenerate He donor scenario.  While subsequent work on double detonations predicted that even these small He shells would adversely affect observations \citep{fhr07,fink10,krom10,sim10,wk11}, more recent multi-dimensional work allowing for post-shock radial expansion in the He layer suggests that He-burning will be truncated before significant production of IGEs \citep{tmb12,holc13a,moll13a,moor13a}.  A large amount of C/O pollution in the He layer, either dredged up from the core or produced during a phase of convective He-burning, may also prevent overproduction of IGEs \citep{krom10,wald11}.

Evolutionary scenarios involving the ignition of He detonations during dynamically unstable He+C/O or C/O+C/O WD mergers have also been studied recently.  In these systems, He is present due to a He WD companion or from the small $10^{-3}-10^{-2} \msol$ He layers that blanket C/O WDs \citep{it85,pakm13a}.  These scenarios include detonations due to the interaction of the direct impact accretion stream with the previously accreted material \citep{guil10,rask12}, contact-induced detonations during the tidal disruption phase of the merger \citep{dan12,pakm12b}, and detonations due to viscous heating of the post-merger configuration \citep{schw12}.  As in the evolutionary channel involving stable mass transfer from a He WD, the amount of He at densities high enough to produce IGEs during a subsequent He detonation in these dynamical scenarios is small.  Thus, these He detonations may also avoid significant contamination of an ensuing SN Ia, especially after accounting for their multi-dimensional nature.

The double WD merger pathway has the additional benefit of yielding several H-rich ejection episodes prior to the dynamically unstable mass transfer and subsequent SN Ia \citep{shen13a}.  These ejection events occur because the H-rich layer surrounding the less massive WD is transferred in a dynamically stable fashion onto the more massive WD and is subsequently blown out of the system in classical nova-like events $300-1500$ yr prior to the SN Ia.  The absorption of the SN light by this previously ejected material yields features that match recent observations of circumstellar material surrounding $10-30\%$ of SNe Ia \citep{pata07,blon09,simo09,ster11,ster13a,fole12a,magu13}.

In summary, multiple evolutionary channels can yield a detonation in a He-rich layer surrounding a degenerate C/O core.  While further work is required to probe the actual ignition of these He detonations, they remain a plausible outcome of mass transfer in some WD binaries.  Throughout the rest of this work, we assume that the He detonation propagates successfully around the entire WD surface and sends a converging shock wave into the core.  The focus of this paper is the convergence of this shock wave and the ignition of the second, C-powered, detonation.


\section{Planar detonation and spherical converging shock analytics}
\label{sec:analytics}

To set the stage for our numerical reactive hydrodynamic simulations, we first consider the simpler problems of planar detonations and non-reactive spherically symmetric imploding shock waves.


\subsection{Chapman-Jouguet results for planar detonations}
\label{sec:CJ}

\begin{table*}
	\begin{center}
	\caption{Chapman-Jouguet detonation speeds for combustion to pure $^{28}$Si}
	\begin{tabular}{|c|c|c|c|c|c|c|}
	\hline
	Composition & $q$ & $\rho_0$ & $c_{s,0}$ &  $\gamma $ & $v_{\rm CJ}$ & $M_{\rm CJ}$\\
	  & $ \left( 10^{17}  {\rm \ erg  \ g}^{-1} \right) $ & $ \left( {\rm g \ cm}^{-3} \right) $ &  $\left( 10^8 \cgsv \right) $ & & $\left( 10^8 \cgsv \right) $ & \\
	\hline
	\hline
	$0.5/0.5$ $^{12}$C$/^{16}$O & 6.0 & $3.2\E{6}$ & $2.69$ & 1.35 & $9.9$ & 3.7 \\
	\hline
	  & & $1.0\E{7}$ & $3.47$ & 1.36 & $10.1$ & 2.9 \\
	\hline
	  & & $3.2\E{7}$ & $4.37$ & 1.38 & $10.4$ & 2.4 \\
	\hline
	$0.7/0.3$ $^{16}$O$/^{20}$Ne & $4.4$ & $3.2\E{6}$ & $2.69$ & 1.37 & $8.8$ & 3.3 \\
	\hline
	  & & $1.0\E{7}$ & $3.47$ & 1.38 & $8.9$ & 2.6 \\
	\hline
	  & & $3.2\E{7}$ & $4.36$ & 1.39 & $9.0$ & 2.1 \\
	\hline
	\end{tabular}
	\label{tab:CJ}
	\end{center}
	Column 1: Initial composition, by mass fraction; Column 2: Specific energy release from converting the initial composition into pure $^{28}$Si; Column 3: Initial density; Column 4: Initial sound speed for an initial temperature of $10^7$ K; Column 5: $\gamma$ at CJ conditions; Column 6: CJ detonation velocity; Column 7: Mach number of CJ detonation with respect to the unburned material.
\end{table*}

The framework used to calculate planar post-detonation conditions is often referred to as the CJ solution, after the original work of \cite{chap99} and \cite{joug05}.  From mass, momentum, and energy conservation, and the assumption that the burned ashes move at the speed of sound in the shock's rest frame, the CJ detonation velocity is $v_{\rm CJ} = \sqrt{ 2 (\gamma^2-1) q }$.  Here $q$ is the energy per mass released from burning the fuel to ash, and the equation of state exponent, $\gamma$, is typically near $1.4$ for our relevant conditions, but must be calculated self-consistently.

For our CJ calculations, we take the end state of burning to be $^{28}$Si because our successful detonations quickly burn the C/O to a state of quasi-nuclear statistical equilibrium (quasi-NSE), consisting of isotopes with binding energies near that of $^{28}$Si.  Given enough time, the material in the propagating detonations will burn all the way to NSE, which involves a mix of IGEs.  However, this occurs on lengthscales and timescales that are orders of magnitude larger than required to burn to quasi-NSE.  Since we are only concerned with small volumes surrounding the focal point, the successes of the detonations in this paper are determined solely by the binding energy released from converting C/O to quasi-NSE.  Table \ref{tab:CJ} shows the CJ results under this assumption for our different initial conditions.


\subsection{Non-reactive spherically symmetric converging shocks}

We now consider the simplified case of spherical imploding shock waves in the absence of chemical reactions.  These types of shocks are referred to as CCW shocks due to the pioneering studies of \cite{ches54}, \cite{chis55,chis57}, and \cite{whit57,whit58}.  Thorough analytic and numeric work on these imploding shock waves has already been performed (e.g., \citealt{gude42,stan60,zr67,ll87,ponc06,klw12}), so we only summarize their results.

The strength of a spherically symmetric shock wave in a converging medium increases as the surface area of an imploding shock front decreases, and decreases as an exploding shock expands.  The Mach number of the shock wave, $M$, scales with the shock's distance from the focal point as $M \propto r^\alpha$ \citep{chis57}, where
\be
	\alpha = - 2 \frac{M^2-1}{\lambda M^2 } , \nonumber
\ee
\be
	\lambda = \lp 2 \sigma + 1 + \frac{1}{M^2} \rp \left[ 1 + \frac{2 \lp 1-\sigma^2 \rp }{\sigma \lp \gamma+1 \rp } \right]  {\rm , \ and} \nonumber
\ee
\be
	\sigma^2 = \frac{ \lp \gamma-1 \rp M^2+2}{ 2 \gamma M^2- \lp \gamma-1 \rp } .
\ee
The equation of state exponents, sometimes referred to as $\Gamma_1$ and $\Gamma_3$, are presumed to be equal, as appropriate for the cases of ideal gas and radiation, and are denoted as $\gamma$.  In the limit of a strong shock with $\gamma = 1.4$, which will be relevant for our future calculations, the Mach number scales as $M \propto r^{-0.39}$.  Because $\alpha$ depends implicitly on $M$, it must be calculated numerically.

These results are derived under the assumption that the evolution of the previously shocked material does not significantly affect the shock properties and that $\gamma$ remains the same before and after the shock.  In spite of these assumptions, the analytic relations compare very well with numerical hydrodynamics results, as we show in the next section.


\section{Numerical reactive hydrodynamics calculations}
\label{sec:numerics}

Because of the non-linearity of nuclear burning, the addition of chemical reactions to the imploding shock formalism complicates the derivation of robust analytic results, although see \cite{klw12} for strong efforts in this direction.  Thus, in this section, we extend the analytic CJ and CCW frameworks with numeric calculations that include nuclear burning.

We utilize the Eulerian hydrodynamics adaptive mesh refinement code, FLASH \citep{fryx00}, which includes the Helmholtz equation of state \citep{ts00b} and a 13-isotope nuclear burning network \citep{timm00c}.  All calculations are performed in one-dimensional spherical symmetry.  Initially, all the material in the computational domain is at rest with a temperature of $10^7$ K.  Material within a sphere of radius $r_0$ has an initial density $\rho_0$.  This sphere is surrounded by a shell with a higher density $\rho_{\rm pert}$ extending to the edge of the domain at a radius of $1.1 r_0$.  The higher density in the shell implies a higher pressure, which creates the inwardly moving shock wave.  An outflow (zero-gradient) boundary condition is implemented at the outer edge of the computational grid.  Since the spatial scale of the C/O detonation initiation site is much less than the pressure scale height and the stellar radius, all of the work in this paper assumes that the unperturbed region has an initially constant density with negligible gravity.  Burning is turned off in zones within shock fronts.  Convergence studies of the effects of the minimum resolution and the ratio of $\rho_{\rm pert}/\rho_0$ were performed and are described Section \ref{sec:convergence}.


\subsection{Purely hydrodynamic converging shocks}
\label{sec:numanalytics}

\begin{figure}
	\plotone{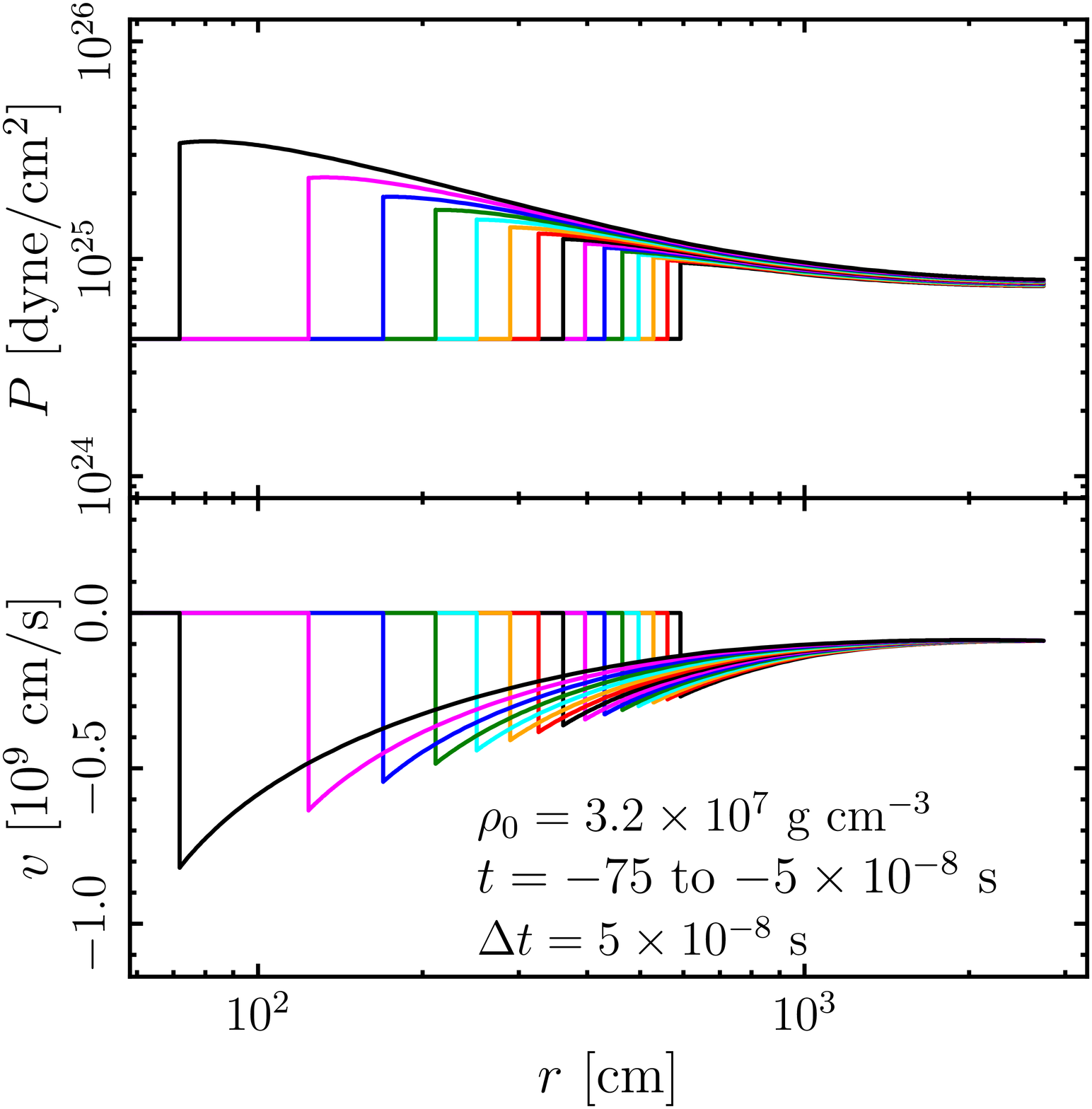}
	\caption{Profiles of pressure (\emph{top panel}) and velocity (\emph{bottom panel}) vs.\ radius from the focal point for an imploding shock wave at snapshots separated by $5\E{-8}$ s, beginning $7.5\E{-7}$ s before the shock wave reaches the focal point.  The initial density is $3.2\E{7} \cgsd$.  Nuclear burning is not included.}
	\label{fig:pvvsr_in_noburn}
\end{figure}

\begin{figure}
	\plotone{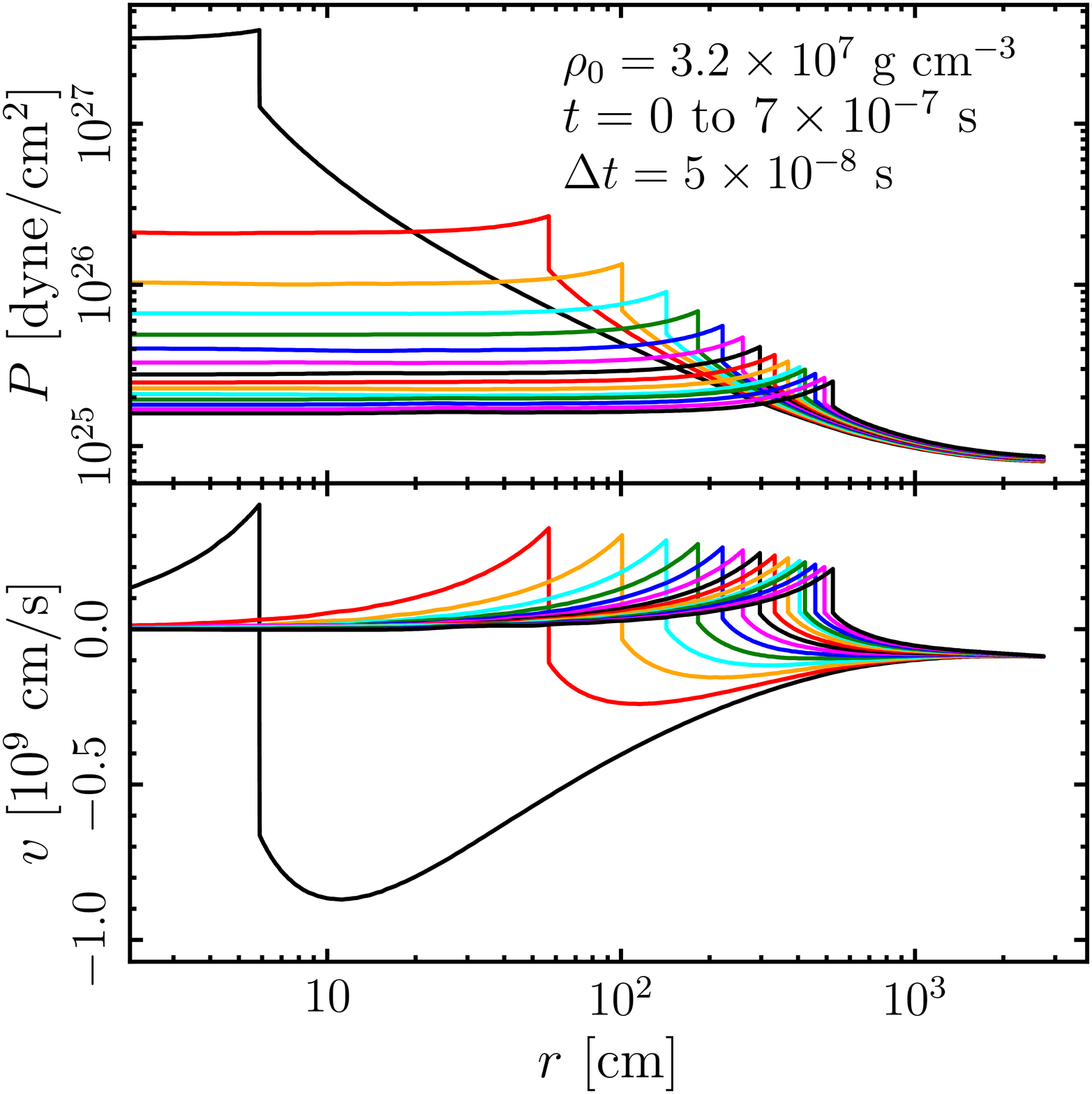}
	\caption{Same as Fig. \ref{fig:pvvsr_in_noburn}, but beginning just after the shock wave has reached the focal point.}
	\label{fig:pvvsr_out_noburn}
\end{figure}

As a first test of the numeric code, we compare the evolution of a purely hydrodynamic (i.e., non-reactive) imploding shock wave to the analytic results from Section \ref{sec:analytics}.  Figure \ref{fig:pvvsr_in_noburn} shows radial pressure (\emph{top panel}) and velocity (\emph{bottom panel}) profiles of an imploding shock wave at 15 snapshots in time.  The initial density is $3.2\E{7} \cgsd$, and the snapshots are separated by $5\E{-8}$ s and begin $7.5\E{-7}$ s before the shock wave has reached the focal point.  Figure \ref{fig:pvvsr_out_noburn} shows the second half of this calculation, beginning just after the shock wave has reached the focal point and reversed its direction.

\begin{figure}
	\plotone{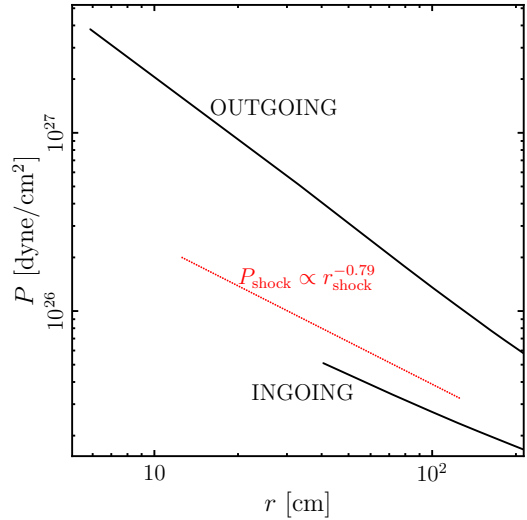}
	\caption{Post-shock pressure vs.\ position of the shock front for the purely hydrodynamic simulation shown in Figs. \ref{fig:pvvsr_in_noburn} and \ref{fig:pvvsr_out_noburn} (\emph{black solid lines}).  The ingoing and outgoing shocks are as labeled.  Also shown is the expected scaling for a strong spherical imploding or exploding shock as given in Section \ref{sec:analytics} (\emph{red dotted line}).}
	\label{fig:lpshockvslrshock}
\end{figure}

Figure \ref{fig:lpshockvslrshock} shows the post-shock pressure versus the position of the shock front for the same calculation shown in Figures \ref{fig:pvvsr_in_noburn} and \ref{fig:pvvsr_out_noburn}.  The relative pressure jump across a shock front is
\be
	\frac{P_1}{P_0} = \frac{2 \gamma M^2 - (\gamma-1)}{\gamma+1} ,
\ee
so the power-law scaling of post-shock pressure with shock radius will be roughly twice as strong as the Mach number's scaling; thus, $P \propto r^{-0.79}$ for a strong shock with $\gamma=1.4$.  This relation is shown as a red dotted line in Figure \ref{fig:lpshockvslrshock}.  Both the imploding and exploding shock fronts follow this scaling fairly well.  The outgoing shock has a higher normalization than the ingoing because it propagates outwards into previously shocked and converging material.

\subsection{Converging shocks with nuclear reactions}
\label{sec:burning}

\begin{figure}
	\plotone{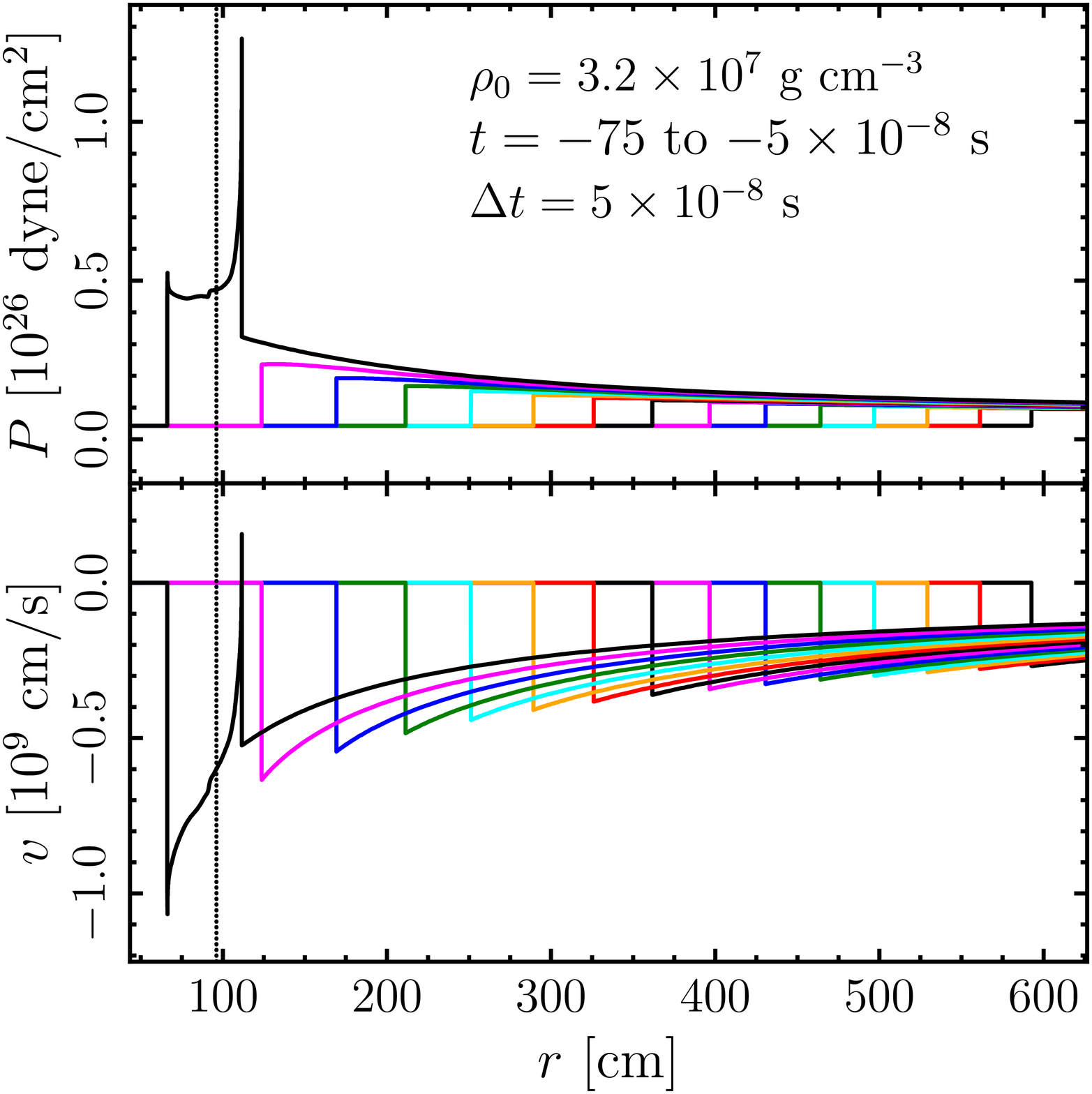}
	\caption{Same as Fig. \ref{fig:pvvsr_in_noburn}, but with nuclear reactions included and with $P$ and $r$ on a linear scale.  The radius where the ingoing shock velocity equals the planar CJ detonation velocity, $r(v_{\rm shock} = v_{\rm CJ}) = 96$ cm, is shown as a dotted line.}
	\label{fig:pvvsr_in_yesburn}
\end{figure}

\begin{figure}
	\plotone{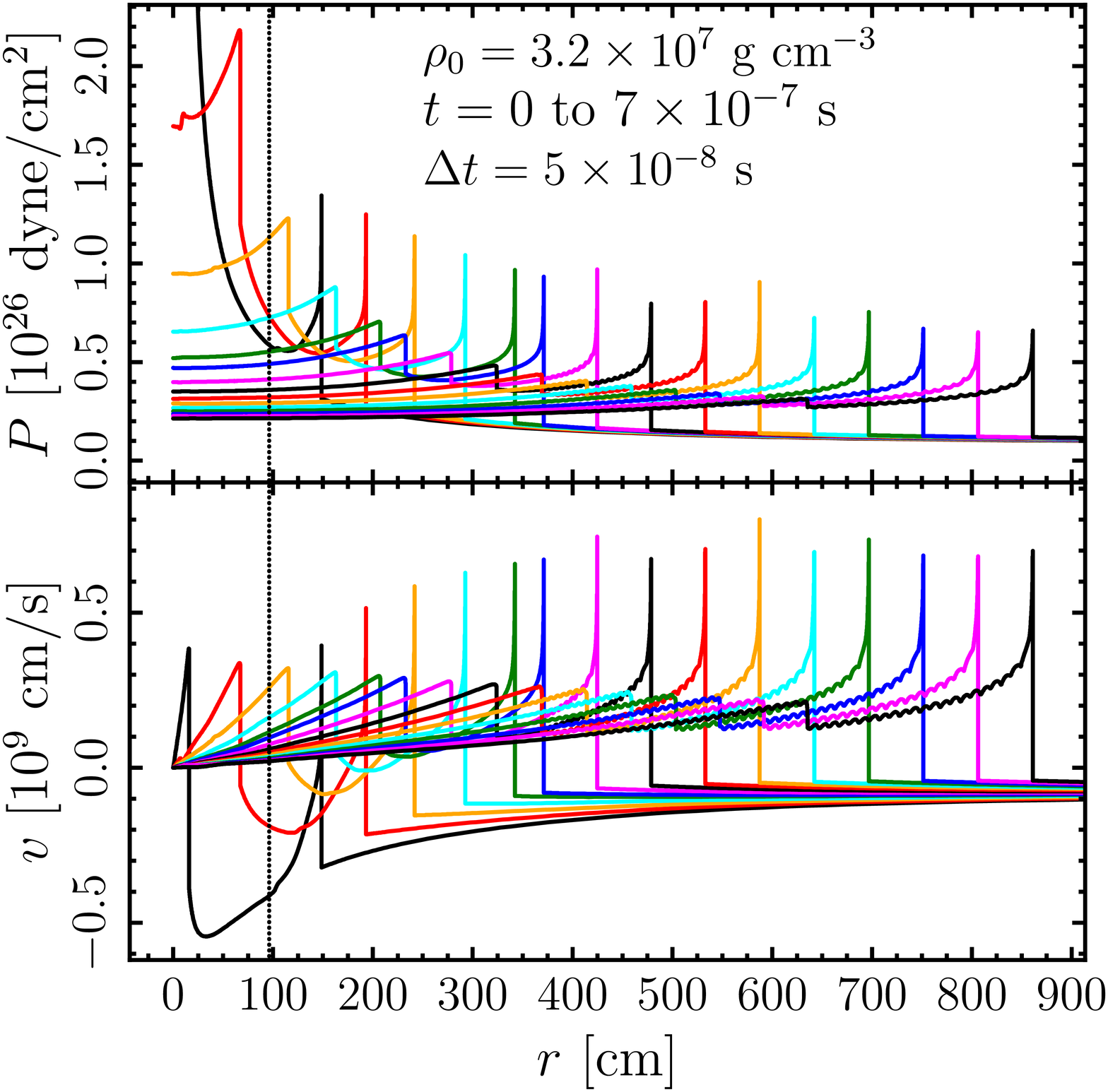}
	\caption{Same as Fig. \ref{fig:pvvsr_in_yesburn}, but beginning just after the shock wave has reached the focal point.}
	\label{fig:pvvsr_out_yesburn}
\end{figure}

When nuclear reactions are included, burning begins as the imploding shock wave approaches the focal point and the post-shock conditions reach burning temperatures and densities.  Figures \ref{fig:pvvsr_in_yesburn} and \ref{fig:pvvsr_out_yesburn} show a calculation with the same initial conditions as the run in Figures \ref{fig:pvvsr_in_noburn} and \ref{fig:pvvsr_out_noburn}, but with nuclear reactions turned on.  As Figure \ref{fig:pvvsr_in_yesburn} demonstrates, nuclear reactions yield an outwardly propagating detonation before the imploding shock wave has reached the focal point.  The outgoing detonation can be seen as a growing spike in the pressure and velocity profiles.

Throughout this paper, we use the radius where the imploding shock velocity equals the planar CJ detonation velocity, $r(v_{\rm shock} = v_{\rm CJ})$, as a proxy for the strength of the initial imploding shock.  A simulation with a larger value of $r(v_{\rm shock} = v_{\rm CJ})$ has a stronger initial shock.  This radius is essentially equivalent to the point at which heating due to nuclear reactions overtakes compressional heating due to the converging shock flow.  For the fiducial calculation shown in Figures \ref{fig:pvvsr_in_yesburn} and \ref{fig:pvvsr_out_yesburn}, this radius is $r(v_{\rm shock} = v_{\rm CJ}) = 96$ cm and is shown as a dotted line.

\begin{figure}
	\plotone{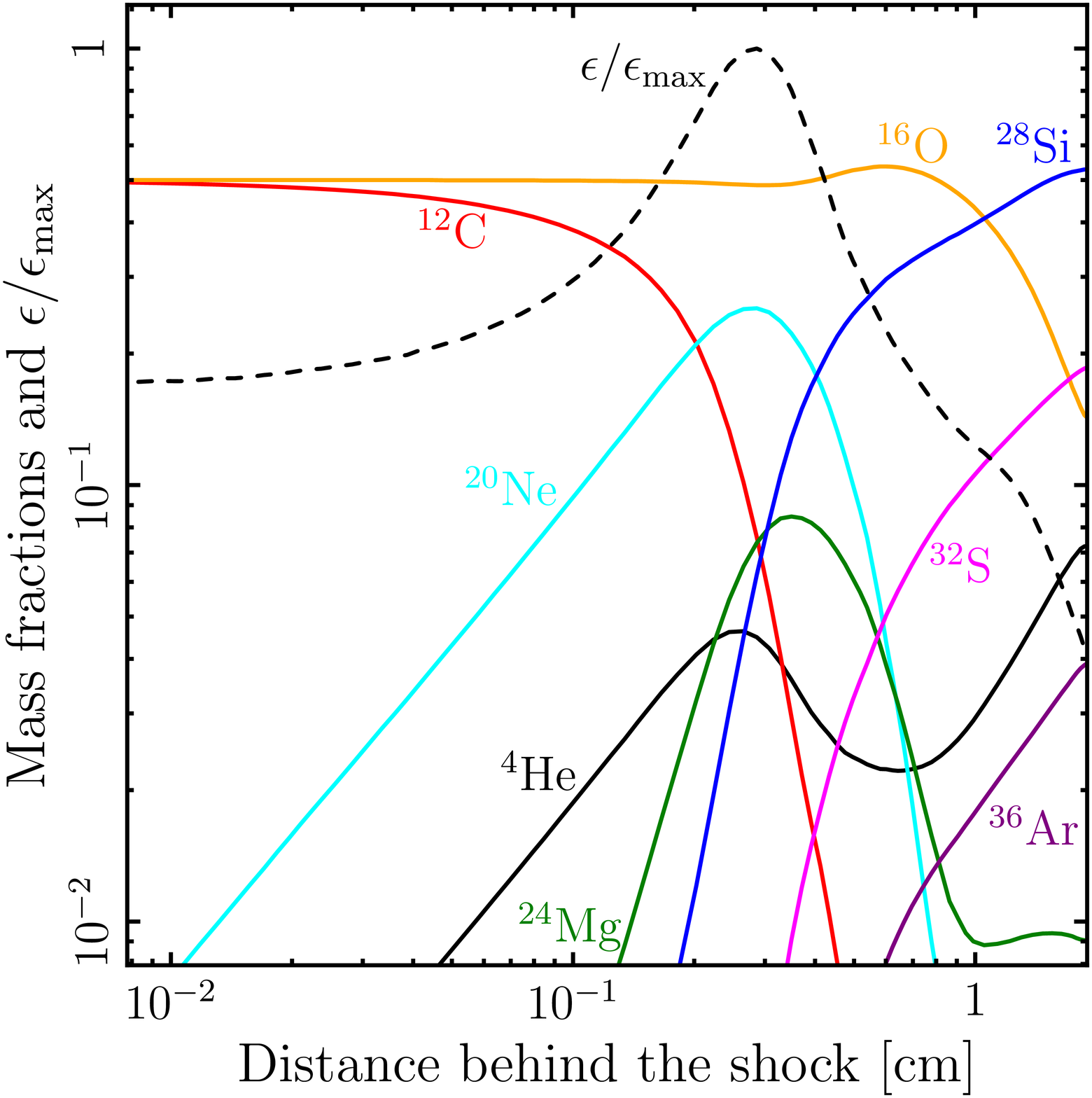}
	\caption{Mass fractions (\emph{solid lines}) and normalized energy generation rate (\emph{dashed line}) vs.\ distance behind the shock for a successfully propagating C detonation.  The shock front is moving at $v= 1.1\E{9} \cgsv$ and is $2.7\E{3}$ cm from the focal point.  The initial density near the focal point was $3.2\E{7} \cgsd$, but the density immediately ahead of the detonation is now $5.9\E{7} \cgsd$ due to the previous passage of the imploding shock wave.}
	\label{fig:lxvslr}
\end{figure}

The spherically symmetric geometry allows us to resolve the very small lengthscales that characterize C detonations at these densities.  Figure \ref{fig:lxvslr} shows the post-shock structure of a successful detonation at the end of the calculations shown in Figures \ref{fig:pvvsr_in_yesburn} and \ref{fig:pvvsr_out_yesburn}.  While the initial density of the material was $3.2\E{7} \cgsd$, the upstream material at this stage of the calculation has a density of $5.9\E{7} \cgsd$ due to the converging shock flow.  Solid lines denote the mass fractions of the most abundant isotopes as labeled, while the dashed line shows the energy generation rate, $\epsilon$, normalized to its maximum value of $4\E{26}$ erg g$^{-1}$ s$^{-1}$.  The shortest detonation lengthscale is associated with the consumption of $^{12}$C ($ 0.2$ cm), followed by the location of the maximum of the energy generation rate ($0.3$ cm), and trailed by the $^{16}$O consumption lengthscale ($ 1.4$ cm).  Note that while the first reactions to take place behind the shock front are $^{12}$C$+^{12}$C self-reactions, the burning material soon approaches quasi-NSE, consisting primarily of $^{28}$Si and $^{32}$S.  Due to the much higher temperatures reached near the focal point, which is $ 2.7\E{3}$ cm farther downstream and not shown in the figure, the composition of the detonation ashes is much closer to full NSE.

\begin{figure}
	\plotone{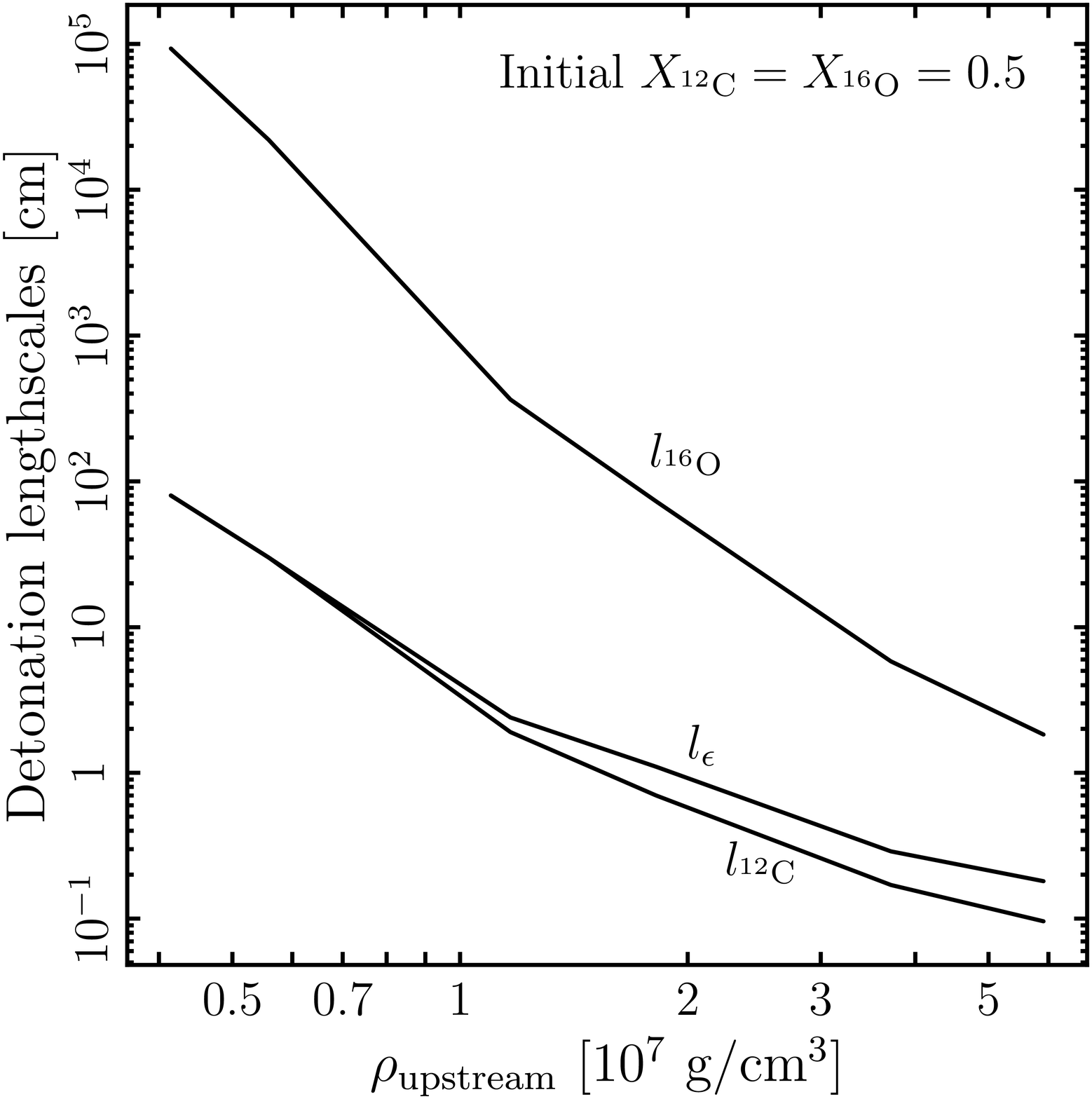}
	\caption{Detonation lengthscales vs.\ upstream density at the end of calculations with successfully propagating C detonations.  Lengthscales are distances between the shock front and the locations of the maximum of the energy generation rate, labeled $l_{\epsilon}$, and the locations where the mass fractions of $^{12}$C and $^{16}$O have been halved, labeled $l_{\rm ^{12}C}$ and $l_{\rm ^{16}O}$, respectively.}
	\label{fig:ZND}
\end{figure}

Figure \ref{fig:ZND} shows a summary of these measures of the detonation lengthscale versus upstream density for calculations with successfully propagating detonations.  The initial mass fractions are $X_{\rm ^{12}C}=X_{\rm ^{16}O}=0.5$.  As in Figure \ref{fig:lxvslr}, the shortest lengthscale is the distance to the location where $X_{\rm ^{12}C}$ has been halved, labeled $l_{\rm ^{12}C}$.  The location of the maximum of the energy generation rate, labeled $l_\epsilon$, is at a distance that is a factor of $1-2$ longer than $l_{\rm ^{12}C}$.  The lengthscale for $^{16}$O to be halved, labeled $l_{\rm ^{16}O}$, is significantly longer than both $l_{\rm ^{12}C}$ and $l_\epsilon$.  The values of these lengthscales are in good agreement with previous work \citep{khok89,game99}.


\subsection{Successful versus unsuccessful detonations}

\begin{figure}
	\plotone{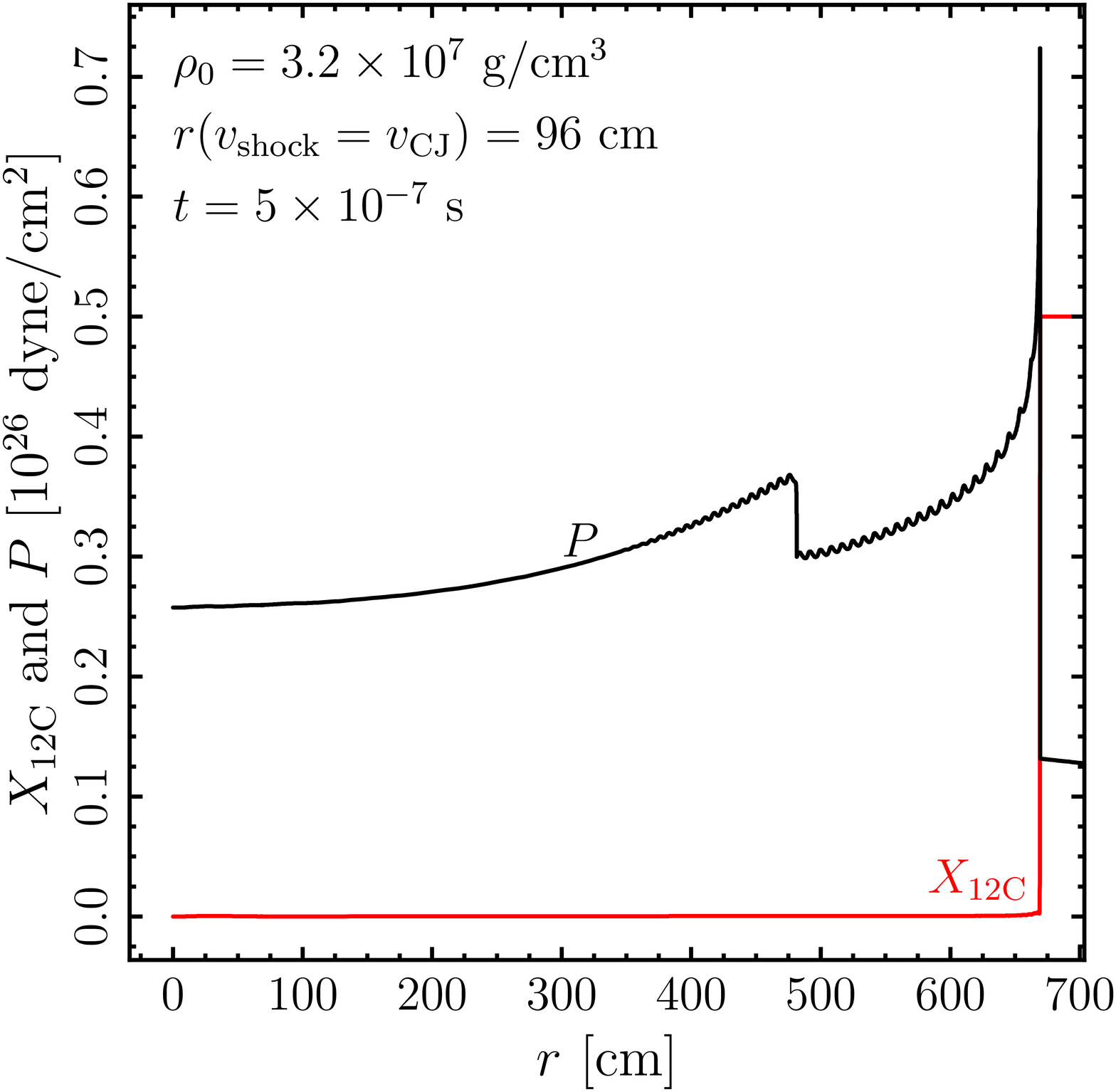}
	\caption{Radial profiles of $P$ (\emph{black line}) and $X_{\rm ^{12}C}$ (\emph{red line}) for a successful detonation $5\E{-7}$ s after the imploding shock wave has reached the focal point.  The initial density at the focal point was $3.2\E{7} \cgsd$, and the initial shock strength implied $r(v_{\rm shock}=v_{\rm CJ})=96$.  The detonation's success can be seen in the coupling of the shock front and the composition discontinuity.}
	\label{fig:pxvsr_yesdet}
\end{figure}

\begin{figure}
	\plotone{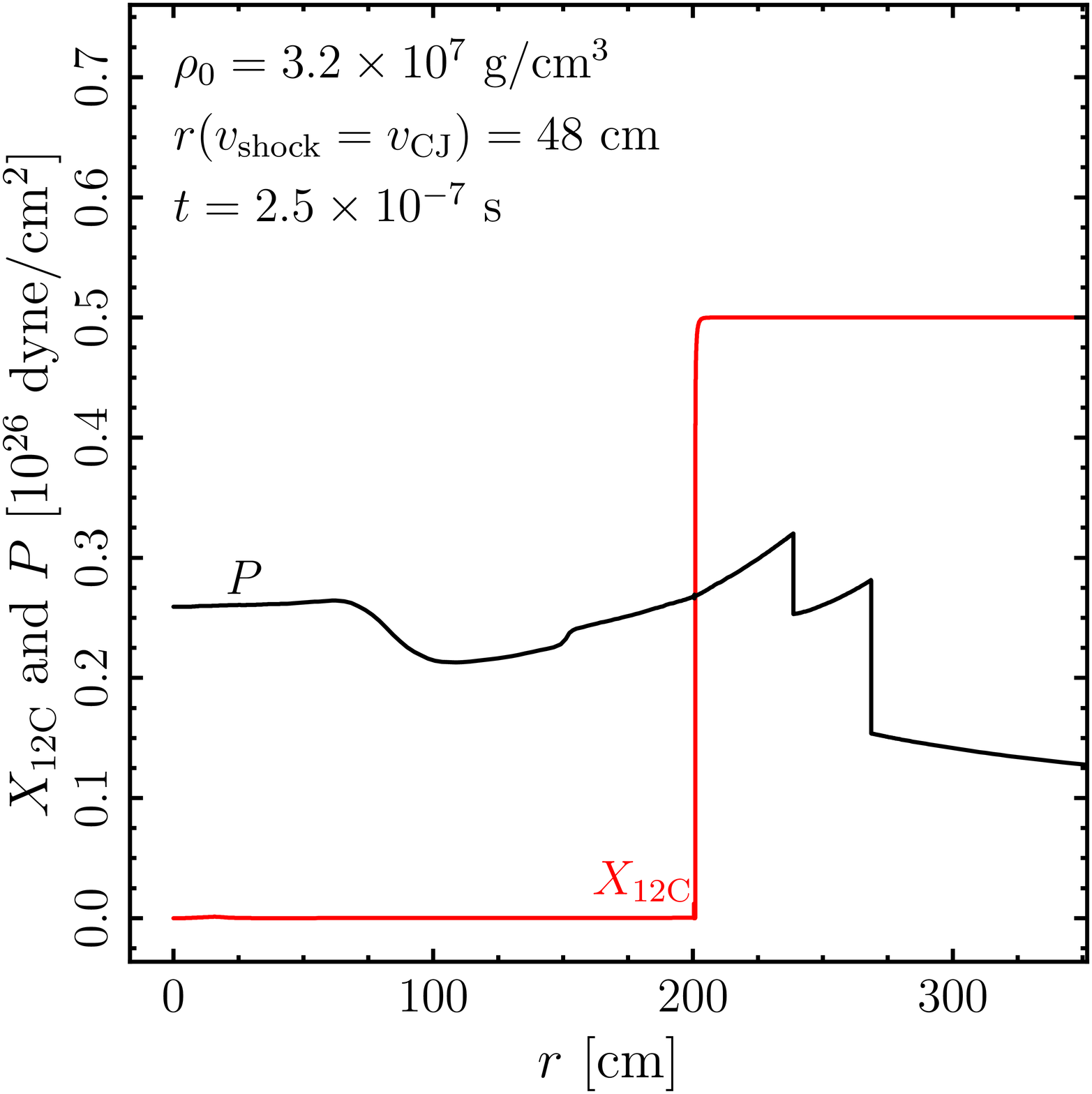}
	\caption{Same as Fig. \ref{fig:pxvsr_yesdet}, but for a weaker initial shock, with $r(v_{\rm shock}=v_{\rm CJ})=48$ cm, at a time $2.5\E{-7}$ after the shock has reached the center.  Due to the weaker initial shock, the detonation is unsuccessful: both outgoing shocks are decoupled from the composition discontinuity, which has ceased moving outwards in a Lagrangian sense.}
	\label{fig:pxvsr_nodet}
\end{figure}

The success of a detonation depends on the initial strength of the imploding shock.  Figures \ref{fig:pxvsr_yesdet} and \ref{fig:pxvsr_nodet} compare a successfully propagating detonation and an unsuccessful detonation, respectively.  Both figures show radial profiles of the pressure (\emph{black lines}), normalized to $10^{26}$ dyne cm$^{-2}$, and of the $^{12}$C mass fraction (\emph{red lines}) at the same relative time in each simulation, measured from the moment when the imploding shock wave reaches the focal point.  Both calculations begin with an initial density of $3.2\E{7} \cgsd$, but the outer boundary for the calculation in Figure \ref{fig:pxvsr_yesdet} is at $2.75\E{3}$ cm, while the simulation in Figure \ref{fig:pxvsr_nodet} has an outer boundary at $1.375\E{3}$ cm.  As a result, the lengthscales and timescales in the successful calculation are twice as large as in the unsuccessful run.  The larger value of $r(v_{\rm shock}=v_{\rm CJ})$ in Figure \ref{fig:pxvsr_yesdet} implies a larger initial shock strength, which is why it forms a successfully propagating detonation, as demonstrated by the superposition of the shock front and the compositional discontinuity.  The simulation with the weaker initial shock fails to yield a detonation, as shown by the lack of coupling of the shock front and the compositional discontinuity, whose velocity in mass space has stalled.

These expanding detonations fail to propagate for two reasons.  First, the velocity is constrained to be zero at the focal point, located behind the outwardly propagating detonation, while for a standard CJ detonation, the downstream ashes move at a finite velocity in the direction of the detonation.  The zero-velocity boundary condition exerts a backwards ``pull'' on the material behind the detonation, preventing it from reaching CJ conditions.  This effect has been demonstrated for outwardly propagating explosions with a zero-velocity boundary condition in planar and spherical geometry by \cite{hc94} and \cite{seit09}, among others.  While a simple estimate of the lengthscale of the critical shocked volume necessary for a subsequent successful detonation might be $\sim l_\epsilon$, the actual critical lengthscale for successful planar C detonations with a zero-velocity boundary condition is a factor of $ 10^3-3\E{4} $ times larger.

The second major effect is due to the spherical curvature of the detonation front.  As the post-shock, but pre-burned, material expands, the density and temperature are reduced and nuclear reactions proceed at a slower pace than for a a planar detonation at the same velocity.  Furthermore, when burning releases most of its energy at some finite distance $\sim l_\epsilon$ behind the shock, it powers a shock front with a larger surface area than in the planar case.  The effect of curvature increases the critical radius by an additional factor of three as compared to $l_\epsilon$, so that the critical radius found by \cite{seit09} was $3\E{3}-10^5$ times larger than the detonation lengthscale, $l_\epsilon$.

However, as we show in the next section, our critical radii are $300-10^4$ larger than the detonation lengthscales for material at the initial density.  This somewhat smaller increase is due to the converging shock flow: the outwardly moving detonation propagates into previously shocked material, which consequently has a shorter detonation lengthscale than it would if it had the initial unshocked density.


\subsection{Critical shock strength for successful detonations and connection to multi-dimensional simulations}
\label{sec:rCJ}

\begin{figure}
	\plotone{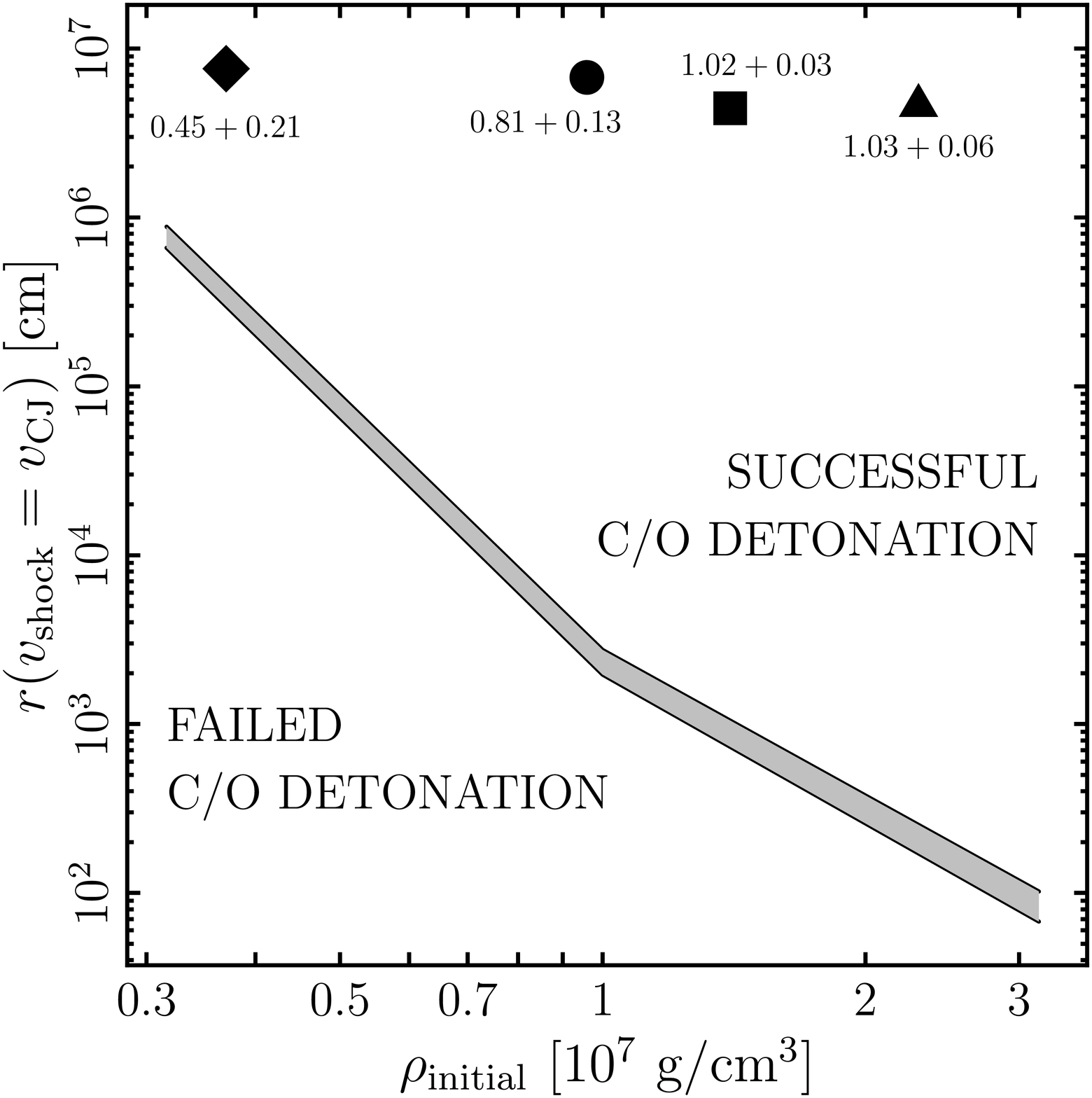}
	\caption{Critical shock strength vs.\ initial density at the focal point necessary for successful detonations.  Spherically symmetric converging shock waves that would reach $v_{\rm CJ}$ at radii larger than the shaded region will lead to successful detonations in C/O material.  Markers represent multi-dimensional full star simulations, labeled with their C/O core $+$ He shell masses.  See text in Section \ref{sec:rCJ} for details.}
	\label{fig:rCJ}
\end{figure}

By varying the size of the computational domain and thus the imploding shock strength, we can map the regions of parameter space that yield successful or unsuccessful detonation ignitions.  Figure \ref{fig:rCJ} shows this dividing line at different initial densities, characterized by the radius at which the imploding shock velocity would equal the CJ detonation velocity if nuclear reactions were neglected.  The critical radius decreases as the initial density increases and ranges from $10^2-10^6$ cm for typical WD densities.  

Also shown in Figure \ref{fig:rCJ} are approximations of radii at which the converging shock wave reaches the CJ velocity from two-dimensional full star hydrodynamic simulations of double detonations.  The diamond, circle, and triangle represent calculations with the PROMETHEUS code \citep{fma89a} of $0.45 \msol$ C/O core $+$ $0.21 \msol$ He shell (Model L of \citealt{sim12}), $0.81+0.13 \msol$ (Model 1 of \citealt{fink10}), and $1.03+0.06 \msol$ (Model 3 of \citealt{fink10}) simulations, as labeled.  The square represents a $1.02 + 0.03 \msol$ simulation (D. Townsley 2013, private communication) calculated with FLASH \citep{fryx00}.  The smallest resolution in these multi-dimensional simulations is $1-3\E{6}$ cm.  The He detonation in each of these simulations is initiated at a point.

Since the converging shock waves in the multi-dimensional simulations are somewhat aspherical, a direct mapping of their shock strengthening to our spherically symmetric calculations requires an approximation.  We first estimate the radius of the imploding shock wave at a given time by calculating the enclosed volume within the shock, $V_{\rm encl}$, and then inferring the spherically averaged radius, $(3 V_{\rm encl}/ 4 \pi)^{1/3}$.  The time evolution of this quantity is used to estimate the shock's inward velocity at two radii, from which the value of the shock strengthening scaling is deduced.  These quantities are then used to estimate the radius at which the imploding shock reaches the CJ velocity, which is always $>10$ km.

The spherically-averaged strength of the imploding shock in these multi-dimensional simulations is many orders of magnitude above our critical values for initial densities $\gtrsim 10^7 \cgsd$, which is the central density of a $0.8 \msol$ WD.  Thus, it appears that propagating He detonations can robustly ignite high-mass C/O cores via converging shock waves, even when the small $0.01-1$ cm C detonation lengthscales are resolved.  However, given the much higher critical $r(v_{\rm shock}=v_{\rm CJ})$ at lower densities, detonations in lower-mass C/O cores are not as certain.  A complete analysis of double detonation ignition necessitates adequately resolved multi-dimensional simulations.


\subsection{Convergence studies}
\label{sec:convergence}

\begin{figure}
	\plotone{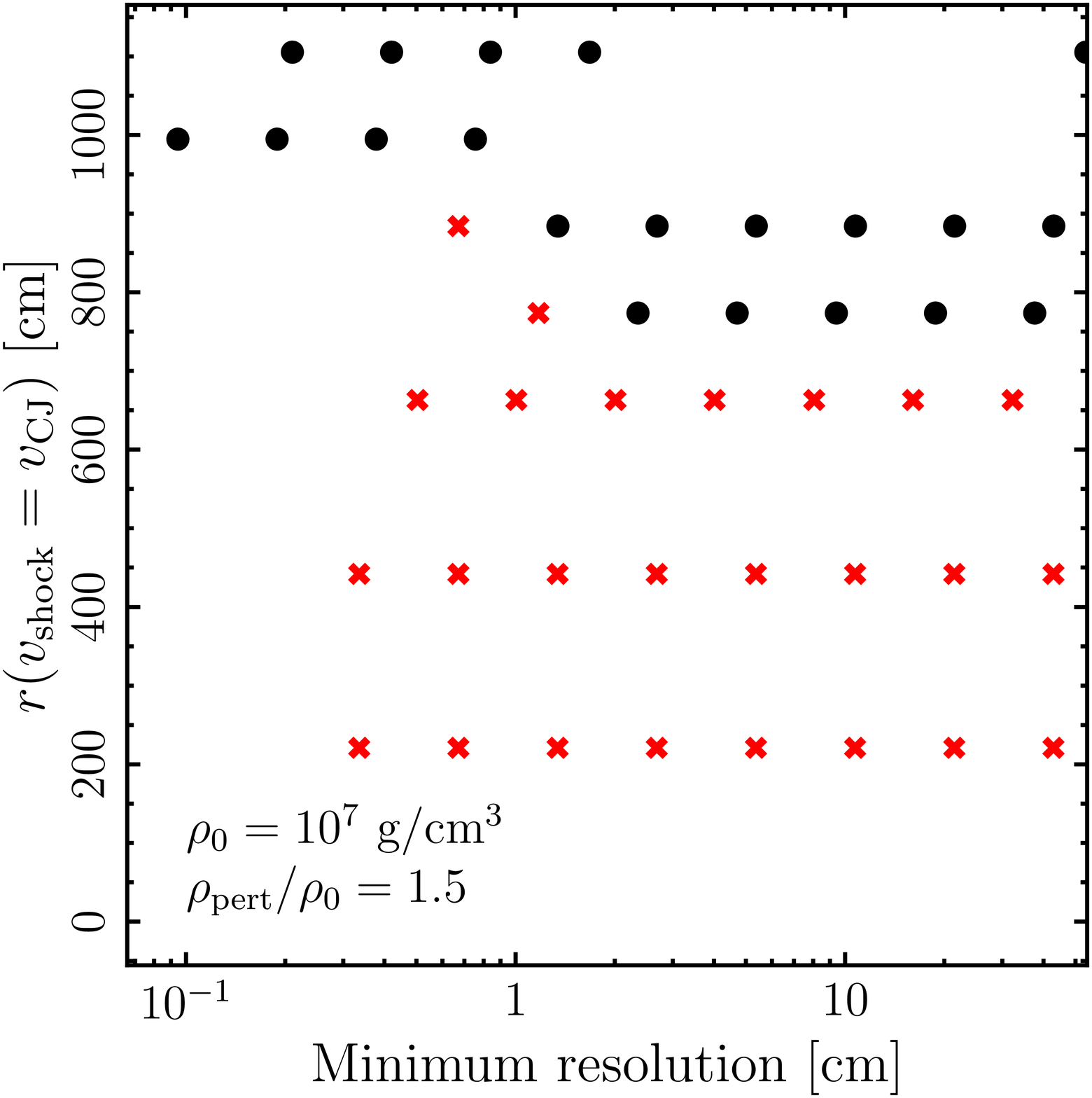}
	\caption{Suite of simulations with different initial imploding shock strengths and minimum resolutions for an initial density of $\rho_0=10^7 \cgsd$ and $\rho_{\rm pert}/\rho_0 = 1.5$.  Black circles (red crosses) demarcate simulations with successful (failed) detonations.  The burning lengthscale is $\sim 0.3$ cm when the detonation develops.  Detonations can propagate successfully in unresolved simulations for initial shocks that are too weak to yield successful detonations in resolved simulations.}
	\label{fig:resstudy}
\end{figure}

\begin{figure}
	\plotone{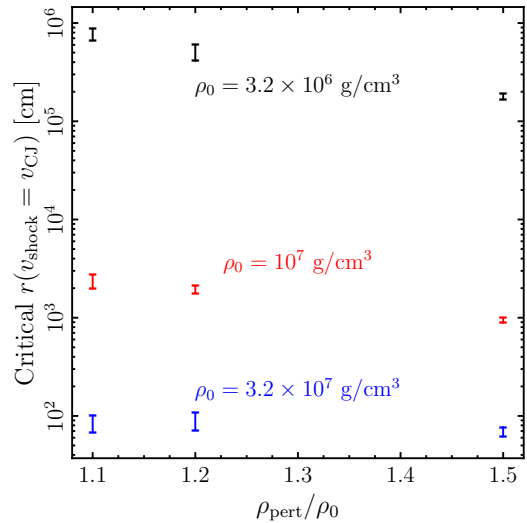}
	\caption{Critical values of $r(v_{\rm shock}=v_{\rm CJ})$ vs.\ $\rho_{\rm pert}/\rho_0$ for $\rho_0 = 3.2\E{6}$, $10^7$, and $3.2\E{7} \cgsd$, as labeled.  A smaller value of $\rho_{\rm pert} / \rho_0$ corresponds to a higher ratio of the size of the computational domain to $r(v_{\rm shock}=v_{\rm CJ})$.}
	\label{fig:vslindensrat}
\end{figure}

To verify that our results are converged, we performed a suite of simulations for each of our three values of $\rho_0$, using three ratios of $\rho_{\rm pert}/\rho_0= 1.1$, $1.2$, and $1.5$ and varying the minimum resolution.  One representative convergence study is shown in Figure \ref{fig:resstudy}, which demonstrates the effect on the critical value for $r(v_{\rm shock}=v_{\rm CJ})$ when the minimum resolution is varied for an initial density of $\rho_0=10^7 \cgsd$ and perturbed density $\rho_{\rm pert} = 1.5 \rho_0$.  Black circles denote simulations with successful detonations; red crosses mark those with failed detonations.  The necessity of resolving the burning lengthscale ($\sim 0.3$ cm for this example) is made clear in simulations with inadequate resolution, which can have successful detonations for imploding shocks that reach the CJ velocity at radii 20\% smaller than the true critical value of 1000 cm.

Figure \ref{fig:vslindensrat} shows the upper and lower bounds on the critical values for $r(v_{\rm shock}=v_{\rm CJ})$ for various ratios of the perturbed density to the initial density, or alternatively, the ratio of $r(v_{\rm shock}=v_{\rm CJ})$ to the size of the computational domain, for our three initial densities as labeled.  Smaller critical values of $r(v_{\rm shock}=v_{\rm CJ})$ are found for simulations with larger $\rho_{\rm pert}/\rho_0$.  This is due to the zero-gradient boundary condition at the outer edge of the grid.  The density and pressure of the once-shocked imploding material at a fixed time should decrease with increasing radius.  The zero-gradient condition thus implies densities and pressures of the inflowing material that are higher than they should be, and when the detonation propagates into this material, the burning lengthscales are correspondingly shorter.  This in turn yields a smaller critical value for $r(v_{\rm shock}=v_{\rm CJ})$.

It is thus important to ensure that the value of $\rho_{\rm pert}/\rho_0$ is low enough to remove the effect of the outer boundary condition.  The critical values of $r(v_{\rm shock}=v_{\rm CJ})$ do indeed converge by $\rho_{\rm pert}/\rho_0=1.1$ for the higher initial densities of $10^7$ and $3.2\E{7} \cgsd$.  In contrast, the derived critical value for the lowest initial density of $3.2\E{6} \cgsd$ is not fully converged even at our lowest value of $\rho_{\rm pert}/\rho_0 = 1.1$, and the computational cost of probing lower values of $\rho_{\rm pert}/\rho_0$ became prohibitive.  However, the exhibited trend implies only a small change to this value for smaller perturbations.  More importantly, the size of the domain for these lowest density simulations approaches the size of a WD, and our assumption of a constant initial density throughout the unperturbed region breaks down.  The critical shock strength for this density should thus only be viewed as an approximation, and very likely a lower limit.


\subsection{Comparison to other work on C detonation initiation}

\cite{klw12} performed an earlier study of detonation initiation from imploding shock waves.  While they apply a more generalized analysis to the problem and focused on reactions with Arrhenius-type temperature dependences, their order-of-magnitude estimate for C/O fuel with $\rho_0 = 10^7 \cgsd$ is $r(v_{\rm shock} = v_{\rm CJ})=10^3$ cm, which is in good agreement with our results.

The spontaneous initiation of spherical C detonations from regions with perturbed temperatures, but unperturbed velocities, has been previously explored by \cite{al94b}, \cite{nw97}, \cite{rwh07a}, and \cite{seit09}.  These authors calculate the critical sizes of hot regions that ignite and yield propagating detonations for various parameterizations of the temperature profile.  Direct comparison of our results to theirs is difficult, as the density ahead of the outwardly propagating detonation in our calculations changes with time, but for a density of $ 3\E{7} \cgsd$ and surrounding temperature of $10^9$ K, \cite{seit09} find critical radii ranging from $3\E{3}-10^5$ cm, depending on the temperature profile of the initiating volume.  The detonation lengthscale at this density is $ \simeq 0.3$ cm, so the ratio of their critical radius to detonation lengthscale is $10^4-3\E{5}$.

In our calculations, for an initial density of $3.2\E{7} \cgsd$, our critical radius is $\simeq 100$ cm, significantly smaller than the lower end of \cite{seit09}'s results.  This is unsurprising because the outward detonation in our case initially propagates into previously shocked material, whose higher density yields smaller detonation lengthscales.  For this initial density of $3.2\E{7} \cgsd$, the newly formed detonation propagates into material with density $> 10^8 \cgsd$ and temperature $\simeq 10^9$ K. Extrapolating \cite{seit09}'s results to these high densities yields a range of critical radii that overlap our value of $100$ cm.


\subsection{O/Ne calculations and the effect of composition}

We have also performed analogous calculations simulating the interiors of O/Ne WDs, with mass fractions of $X_{\rm ^{16}O}=0.7$ and $X_{\rm ^{20}Ne}=0.3$.  However, the increased Coulomb barrier for O-burning results in a much longer detonation lengthscale, which is even larger by a factor of $\sim 10^4$ than the O-consumption lengthscale in a propagating C detonation shown in Figure \ref{fig:ZND}, because O-burning is enhanced in that case by the presence of $^4$He nuclei liberated during C-burning.  Thus, much higher imploding shock strengths and larger simulation volumes are required to achieve a successful detonation, which makes resolution of the detonation structure difficult.  As a result, none of our O/Ne runs that spatially resolved the burning lengthscales yielded a successfully propagating detonation.

\cite{seit09} explored the effect of composition on the spontaneous initiation of detonations and found that decreasing the initial carbon mass fraction from $X_{\rm ^{12}C} = 0.5$ to $X_{\rm ^{12}C}=0.3$ for one of their simulations increased the critical radius for detonation ignition by a factor of $10-100$.  Extrapolating their results to $X_{\rm ^{12}C}=0$ suggests critical radii that are $5\E{4}-10^8$ times larger than for the case with $X_{\rm ^{12}C}=0.5$, and thus it is unsurprising that we have not resolved a successful O/Ne detonation in our calculations.

If the critical radii for O/Ne detonations are as much as $10^6$ times larger than for C/O detonations, Figure \ref{fig:rCJ} suggests that double detonations do not occur if the WD core is C-deficient.  This may explain why detonations of O/Ne WDs, which have masses $\geq 1.2 \msol$ and would yield overluminous SNe Ia with relatively fast light curve evolution, have not been observed \citep{sim10}.


\section{Conclusions}
\label{sec:conc}

In this paper, we have performed numerical calculations that spatially resolve the ignition of the core C detonation in the double detonation scenario.  We have calculated the minimum inward shock strength necessary to achieve a successful outwardly propagating detonation and found that ignition in high-mass C/O cores is plausible if a He shell detonation occurs.  However, O/Ne cores and low-mass C/O cores are harder to ignite, and converging shock waves in such WDs may fail to detonate.

Systems for which the converging shock wave is too weak to initiate a core detonation, either because of low densities or low C abundance, will not lead to SNe Ia.  Since only a small volume near the focal point is heated significantly, this material will just expand, rise buoyantly, and redistribute its entropy without leading to sustained convection or the birth of a deflagration.  However, the radioactive decay of the He detonation ashes in such systems will yield a faint and rapidly evolving ``.Ia'' supernova \citep{bild07,sb09b,shen10,wald11}.

The possibility certainly remains that non-double detonation progenitor channels succeed and contribute to the observed SN Ia population.  For example, the growing class of SNe Ia that exhibit strong interaction with nearby H-rich circumstellar material may in fact be due to the single degenerate scenario \citep{hamu03,alde06,dild12,silv13a,silv13b}.  However, such SNe Ia are relatively rare.  Our work, which puts the success of the core ignition on firmer theoretical ground, makes the growing evidence that double detonations provide a dominant fraction of SNe Ia even more attractive.

While our one-dimensional calculations suggest that double detonations are quite plausible for high-mass C/O WDs, multi-dimensional simulations are still necessary to ensure the robustness of the ignition mechanism, especially with respect to the asphericity of the actual converging shock waves seen in full star simulations.  One possible issue is detonation instability, as seen in the case of imploding detonations \citep{do92,od94}.  A related complication is the multi-dimensional cellular structure of detonations \citep{timm00b}, which can increase their burning lengthscales.  Furthermore, the WD core will be rotating; we expect that as long as the rotation speeds are very subsonic, they will not affect the propagation of the shock waves, but this requires explicit confirmation.  The resolution of these issues awaits future multi-dimensional studies.


\acknowledgments

We thank the anonymous referee for their comments and for motivating us to perform more rigorous convergence studies.  We thank Dan Meiron for early discussions and for referring us to relevant literature.  We are also grateful to Bill Paxton, Josiah Schwab, and Frank Timmes for providing and assisting with code, and to Michael Fink and Dean Townsley for contributing their data.  We thank them, James Guillochon, Doron Kushnir, Ian Parrish, Alexei Poludnenko, Eliot Quataert, Enrico Ramirez-Ruiz, and Ivo Seitenzahl for discussions.  This work was supported by the National Science Foundation under grants PHY 11-25915 and AST 11-09174.  KJS is supported by NASA through Einstein Postdoctoral Fellowship grant number PF1-120088 awarded by the Chandra X-ray Center, which is operated by the Smithsonian Astrophysical Observatory for NASA under contract NAS8-03060.



\end{document}